\documentclass[a4paper, 11pt]{article}
\pdfoutput=1
\usepackage{jcappub}
\notoc{}

\usepackage{feynmp-auto}
\makeatletter
\let\ginnatwidth\Gin@nat@width
\let\ginnatheight\Gin@nat@height
\makeatother
\setkeys{Gin}{width=\linewidth,totalheight=\textheight,keepaspectratio}
\newenvironment{feynmandiagram}[1][]{\setkeys{Gin}{width=\ginnatwidth,totalheight=\ginnatheight}\begin{fmffile}{#1}
\begin{fmfgraph*}(100,70)\fmfpen{thick}}{\end{fmfgraph*}\end{fmffile}\setkeys{Gin}{width=\linewidth,totalheight=\textheight,keepaspectratio}}

\usepackage{graphicx}
\usepackage{booktabs}
\usepackage{verbatim}
\usepackage{xspace}
\usepackage{hyperref}
\usepackage{multirow}
\usepackage{placeins}
\usepackage{array}
\usepackage{subcaption}

\usepackage{lineno}

\setkeys{Gin}{width=\linewidth,totalheight=\textheight,keepaspectratio}

\newcommand{\MET}{\ensuremath{E_T^\mathrm{miss}}\xspace}
\newcommand{\met}{\MET}

\newcommand{\mDM}{\ensuremath{M_{\chi}}\xspace}
\newcommand{\mdm}{\ensuremath{M_{\chi}}\xspace}

\newcommand{\mdmprime}{\ensuremath{M^\prime_{\chi}}\xspace}

\newcommand{\mMed}{\ensuremath{M_{\rm{med}}}\xspace}

\newcommand{\mMedprime}{\ensuremath{M^\prime_{\rm{med}}}\xspace}

\newcommand{\gDM}{\ensuremath{g_{\chi}}\xspace}
\newcommand{\gq}{\ensuremath{g_q}\xspace}
\newcommand{\gl}{\ensuremath{g_l}\xspace}
\newcommand{\gdm}{\gDM}
\newcommand{\ifb}{\ensuremath{\rm{fb}^{-1}}\xspace}

\newcommand{\metplusx}{\ensuremath{\MET+X}\xspace}

\definecolor{cerulean}{RGB}{44,150,207}

\def\be   {\begin{equation}}   \def\ee   {\end{equation}}
\def\ba   {\begin{array}}      \def\ea   {\end{array}}
\def\bea  {\begin{eqnarray}}   \def\eea  {\end{eqnarray}}
\def\bean {\begin{eqnarray*}}  \def\eean {\end{eqnarray*}}

\allowdisplaybreaks

\RequirePackage[normalem]{ulem} 
\RequirePackage{color}\definecolor{RED}{rgb}{1,0,0}\definecolor{BLUE}{rgb}{0,0,1} 

\title{Displaying dark matter constraints from colliders with varying simplified model parameters}

\author[1]{Andreas~Albert,}
\affiliation[1]{Boston University, Dept.\ of Physics, 590 Commonwealth Avenue, Boston, \\ MA 02215, USA}

\author[2]{Antonio~Boveia,}
\affiliation[2]{Ohio State University and Center for Cosmology and Astroparticle Physics, \\
191 W. Woodruff Avenue Columbus, OH 43210, USA}

\author[3]{Oleg~Brandt,}
\affiliation[3]{Cavendish Laboratory, JJ Thomson Avenue, Cambridge CB3 0HE, UK}

\author[4]{Eric~Corrigan,}
\affiliation[4]{Fysiska institutionen, Lunds universitet, Professorsgatan 1, Lund, Sweden}

\author[1]{Zeynep~Demiragli,} 

\author[4]{Caterina~Doglioni,}

\author[5]{Etienne~Dreyer,}
\affiliation[5]{Weizmann Institute of Science, Herzl St 234, Rehovot, Israel}

\author[6]{Boyu~Gao,}
\affiliation[6]{Duke University, Durham, NC 27708, USA}

\author[4]{Josh~Greaves,}

\author[7]{Ulrich~Haisch,}
\affiliation[7]{Max Planck Institute for Physics, F{\"o}hringer Ring 6,  80805 M{\"u}nchen, Germany}

\author[8]{Philip~Harris,}
\affiliation[8]{Massachusetts Institute of Technology, 77 Massachusetts Avenue, Cambridge, MA, USA}

\author[9]{Greg~Landsberg,}
\affiliation[9]{Brown University, Dept.\ of Physics, 182 Hope St, Providence, RI 02912, USA}

\author[10]{Alexander~Moreno,}
\affiliation[10]{Universidad Antonio Nari\~{n}o, Bogotá, Cundinamarca, Colombia}

\author[11]{Katherine~Pachal,}
\affiliation[11]{TRIUMF, 4004 Wesbrook Mall, Vancouver, BC V6T 2A3, Canada}

\author[12]{Priscilla~Pani,}
\affiliation[12]{DESY Zeuthen, Platanenallee 6, 15738 Zeuthen, Germany}

\author[13]{Federica~Piazza,}
\affiliation[13]{INFN Milano, Via G. Celoria, 16 I - 20133 Milano, Italy}

\author[14]{Tim~M.~P.~Tait,}
\affiliation[14]{University of California Irvine, Irvine, CA 92697, USA}

\author[9]{David~Yu,}

\author[15]{Felix~Yu,}
\affiliation[15]{Institut für Physik WA THEP, Johannes Gutenberg-Universität Mainz, Staudingerweg 7, 55128 Mainz, Germany}

\author[16]{Lian-Tao~Wang}
\affiliation[16]{Department of Physics, University of Chicago, Chicago, IL. 60637 USA}

\emailAdd{pcharris@mit.edu}
\emailAdd{kpachal@triumf.ca}

\abstract{
The search for dark matter is one of the main science drivers of the particle and astroparticle physics communities.  Determining the nature of dark matter will require a broad approach, with a range of experiments pursuing different experimental hypotheses. Within this search program, collider experiments provide insights on dark matter which are complementary to direct/indirect detection experiments and to astrophysical evidence.
To compare results from a wide variety of experiments, a common theoretical framework is required. The ATLAS and CMS experiments have adopted a set of simplified models which introduce two new particles, a dark matter particle and a mediator, and whose interaction strengths are set by the couplings of the mediator. 

So far, the presentation of LHC and future hadron collider results has focused on four benchmark scenarios with specific coupling values within these simplified models. In this work, we describe ways to extend those four benchmark scenarios to arbitrary couplings, and release the corresponding code for use in further studies. This will allow for more straightforward comparison of collider searches to accelerator experiments that are sensitive to smaller couplings, such as those for the US Community Study on the Future of Particle Physics (Snowmass~2021)~\cite{snowmass21}, and will give a more complete picture of the coupling dependence of dark matter collider searches when compared to direct and indirect detection searches. By using semi-analytical methods to rescale collider limits, we drastically reduce the computing resources needed relative to traditional approaches based on the generation of additional simulated signal samples. 
} 

\begin{document}

\maketitle



\section{Introduction}
\label{sec:introduction}

The nature of dark matter is one of the most compelling open questions confronting physics today. Across the particle and astroparticle physics communities, a wide range of experiments are attempting to observe and characterise dark matter (DM) and understand what its relationship to the Standard Model (SM) may be. The range of possibilities for the nature of DM and its interactions is vast (see ~\cite{doi:10.1146/annurev-astro-082708-101659} for a review) and as a result, the experimental scope is similarly enormous and the approaches taken highly varied. One important experimental direction is the search for DM production at colliders~\cite{Kahlhoefer:2017dnp,doi:10.1146/annurev-nucl-101917-021008}.

To understand how all the different experiments across different areas of physics complement one another in their collective search for DM, a common set of benchmarks for comparing results is required. The complementarity between experimental frontiers is both highly important and difficult to capture. Indirect detection, direct detection, and collider experiments each have unique strengths in particular areas of the space of possible DM masses and models. To fully explore this space, understand what is covered experimentally, and identify gaps in the broader experimental program, common frameworks have to be defined in which exclusion limits from current and future experiments can be visualised together. This contextualisation requires that the models in which the various experiments interpret their results can be meaningfully compared.

The ATLAS and CMS collaborations produce a vast array of searches which can be interpreted in DM contexts, including final states that correspond to supersymmetric, 2HDM+a, and complicated dark sector models. These specific interpretations make powerful statements about the theories in question but are difficult to compare to one another and to results from other fields. To provide a common framework that would allow for broader interpretation of ATLAS and CMS DM results, the ATLAS/CMS Dark Matter Forum established a specific set of simplified models which the two collaborations have used as the basis for many of their DM searches throughout the LHC Run 2~\cite{ABERCROMBIE2020100371}. In addition to the simplified models themselves, a smaller set of specific benchmarks with fixed coupling values was established, allowing for clear comparisons between and within the LHC experiments~\cite{BOVEIA2020100365,ALBERT2019100377,ATL-PHYS-PUB-2020-021,CMSSummary}.

There are contexts, however, in which these benchmark scenarios are not as flexible as an optimal comparison would require. For example, the HL-LHC and future colliders should be able to probe smaller coupling values than the LHC, low-mass experiments often present results for dark photon models that don't align clearly with the current LHC DM benchmarks, and LHC limits reframed in terms of DM-nucleon interaction cross section can vary with coupling choices in a way which is not fully captured by the existing benchmarks alone.

Historically, reinterpreting ATLAS and CMS analysis limits for a new set of couplings within the simplified model framework required generating new Monte Carlo signal simulations and re-running the statistical analysis. This made testing more than a handful of scenarios prohibitive. But for two signals with the same physical signature in the detector and differing only in cross section, the limit set on one signal can be reinterpreted as a limit on the other by direct rescaling, as long as the ratio between the cross sections of the two signals is known. Having reliable analytical approximations for the cross sections of the simplified models relative to each other will then enable analyses to begin from a single limit in one set of couplings and rescale to any other, as long as the original result includes sufficient information. 

This paper presents analytical and semi-analytical methods enabling rescaling for two of the standard DM simplified models used by ATLAS and CMS with reasonable accuracy and without resorting to generating events. The rescaling formulas are valid for any hadron collider and are based on leading order cross sections for the production of the most important final states for these models. The Python package developed to streamline the application of these techniques can be found in GitHub \href{https://github.com/LHC-DMWG/DMWG-couplingScan-code}{here}. The present version of the code and this whitepaper is intended to support Snowmass~2021~\cite{snowmass21}. We will document the full code as part of a future version of this whitepaper that we plan to publicly release within the LHC Dark Matter Working Group. 

To briefly summarise what is not included in this framework: rescaling certain of the LHC simplified models is not currently supported (see Section~\ref{sec:models}), limit rescaling for lepton colliders is not currently supported, and both analysis acceptances and $k$-factors are treated as constant by the rescaling (see Section~\ref{sec:assumptions}). The first two elements can be added in the future by extensions to the current work if there is sufficient interest from the community. The third, treatment of acceptances and $k$-factors, must be considered on an individual basis by users of the framework.

Additionally, this paper uses a similar analytical approach to determine the minimum allowed coupling values within the simplified models for which the resulting scenario is compatible with the relic abundance of DM. The results provided make it possible to compare the non-excluded points in a mass-mass plane for any set of couplings to the minimum allowed coupling compatible with the relic density in those areas and draw a conclusion about the relevance of those regions. They enable a physically realistic interpretation of the simplified models and illuminate their most interesting regions.

\section{Models considered}
\label{sec:models}

The simplified models considered in this study are detailed extensively in Refs.~\cite{ABERCROMBIE2020100371,ALBERT2019100377}. Each simplified model adds two new particles beyond the Standard Model: a Dirac fermion dark matter particle $\chi$ and an $s$-channel mediator particle. This mediator can be spin-1, in which case its couplings can be either vector or axial-vector in nature, or spin-0, in which case its couplings can be scalar or pseudo-scalar. In this paper, we will refer to these four different mediator types as different models.

Each model introduces two types of new vertex: a vertex between the mediator and a pair of DM particles, and a vertex between the mediator and a pair of SM fermions. There is no interaction between the mediator and SM bosons, and no interaction between the DM particle and any particle other than the mediator. The strength of the interactions at each vertex is governed by a coupling $g$. ATLAS and CMS treat the mediator-SM interactions as having a single universal coupling strength \gq for all mediator-quark interactions and a single universal coupling strength \gl for all mediator-lepton interactions, including neutrinos. Mediators with smaller masses will be kinematically forbidden from decaying into heavy quarks or leptons, but the coupling is still taken to be constant for all fermion generations. For vector and axial-vector mediators, this leads to universal quark and lepton couplings. For scalar and pseudoscalar mediators, the fermion couplings are proportional to the SM Higgs couplings, where the overall relative strengths of the quark and lepton coupling types are scaled by \gq and \gl. No Higgs mixing is included in the spin-0 simplified models, and no mixing with the Z-boson is included in the spin-1 models. These choices mean that the four simplified models are in fact fully defined by five free parameters: the mediator mass \mMed, the DM particle mass \mDM, the coupling strength between the mediator and DM particles \gdm, and the two mediator-SM universal coupling strengths \gq and \gl. In this paper, we will refer to different sets of coupling values within the same model  as different scenarios.

\begin{figure}[h!]
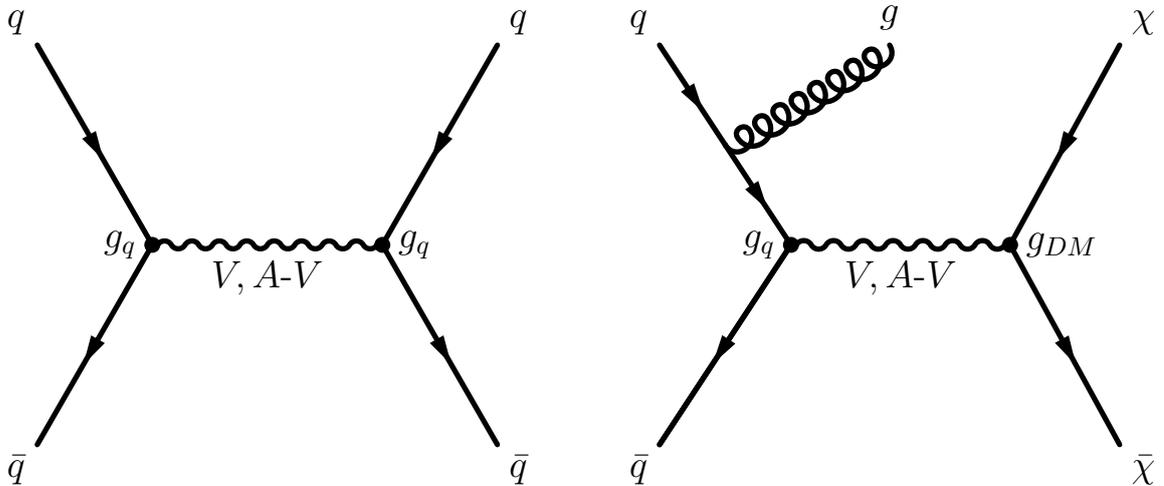

  \begin{center}
  \unitlength=0.005\textwidth
  \vspace{\baselineskip}
  \begin{subfigure}[b]{0.45\textwidth}    
  \begin{feynmandiagram}[av-dijet]
    \fmfleft{i1,i2}
    \fmfright{o1,o2}
    \fmf{wiggly,tension=0.6,label={\Large $V,,A\text{-}V$}}{v1,v2}
    \fmf{fermion,tension=0.6}{o2,v2,o1}
    \fmf{fermion,tension=0.6}{i2,v1,i1}
    \fmfdot{v1,v2}
    \fmflabel{\Large ${g_q}$}{v1}
    \fmflabel{\Large ${g_q}$}{v2}
    \fmflabel{\Large ${\bar{q}}$}{i1}
    \fmflabel{\Large ${q}$}{i2}
    \fmflabel{\Large ${\bar{q}}$}{o1}
    \fmflabel{\Large ${q}$}{o2}
  \end{feynmandiagram}
  \end{subfigure}
  \hfill
  \begin{subfigure}[b]{0.45\textwidth}    
  \begin{feynmandiagram}[av-monojet]
    \fmfleft{i1,i2}
    \fmfright{o1,o2}
    \fmftop{isr}
    \fmfbottom{pisr}
    \fmf{wiggly,tension=0.6,label={\Large $V,,A\text{-}V$}}{v1,v2}
    \fmf{fermion,tension=0.6}{o2,v2,o1}
    \fmf{fermion}{i2,visr,v1}
    \fmf{plain}{v1,pvisr,i1}
    \fmf{fermion,tension=0}{v1,i1}
    \fmfdot{v1,v2}
    \fmflabel{\Large ${g_q}$}{v1}
    \fmflabel{\Large ${g_{DM}}$}{v2}
    \fmflabel{\Large ${\bar{q}}$}{i1}
    \fmflabel{\Large ${q}$}{i2}
    \fmflabel{\Large ${\bar{\chi}}$}{o1}
    \fmflabel{\Large ${\chi}$}{o2}
    \fmf{gluon,tension=0}{visr,isr}
    \fmf{phantom,tension=0}{pvisr,pisr}
    \fmflabel{\Large ${g}$}{isr}
  \end{feynmandiagram}
  \end{subfigure}  
\vspace{\baselineskip}  
\caption[\baselineskip]{Representative Feynman diagrams for leading-order processes with a vector or axial-vector $s$-channel mediator. The visible decay mode (left) uses a dijet final state as an example. The invisible decay mode (right) uses a monojet final state as an example. }
\label{fig:diagrams}
  \vspace{0.5\baselineskip}
\end{center}
\end{figure}

The rescaling methods presented here are computed using leading-order approximations for the cross sections of the key processes. For vector and axial-vector mediators, the relevant processes are resonant SM particle pair production, with or without accompanying initial state radiation (ISR), and invisible decay of a mediator in the presence of ISR leading to a final state of a single SM particle and significant missing energy. Examples of these processes, a dijet and a monojet scenario, are illustrated in Fig.~\ref{fig:diagrams}. Decays of the mediator to leptons, or the presence of an ISR particle (jet, photon, etc), also provide visible signatures with significant exclusion power; mono-photon or other mono-boson analyses are also relevant invisible signatures. However, in all such cases, the change in cross section due to a variation in the DM model or scenario studied can be estimated from the core $qq\rightarrow \text{med} \rightarrow f \bar{f}$ or $qq\rightarrow \text{med} \rightarrow \chi \bar{\chi}$ process; the effect of an ISR particle cancels out in the cross section ratio used for rescaling.

For scalar and pseudoscalar mediators, the leading order DM or SM production diagrams are significantly more complex and additional signatures like $t\bar{t}+\MET$ become important. For this reason, although similar rescaling formulas could be computed for the scalar and pseudoscalar models, they are beyond the scope of the present paper. Rescaling options presented here will therefore focus only on the vector and axial-vector models. Any readers interested in using or helping to develop scalar/pseudoscalar rescaling are invited to contact the authors. Relic density calculations are available for all four models.

\section{Assumptions and caveats}
\label{sec:assumptions}

Although the rescaling formulae are based on the leading order cross section calculations for the visible and invisible final states, recent experimental results are typically reported at next-to-leading order (NLO). NLO accuracy can be reasonably approximated in many cases by multiplying an original exclusion limit calculated at NLO by the LO/LO cross section scale factor computed using the methods here. The assumption involved is that the $k$-factors are unchanged between the start and end points of the rescaling. For example, for the same (\mMed, \mDM) point in an axial model with couplings $g_{a,b,c}$ being rescaled to a vector model with couplings $g_{e,f,g}$, as long as the $k$-factors at these points are the same in the two models and scenarios, a rescaled NLO limit at (\mMed, \mDM) will be NLO also. Similarly, if a visible final state limit defined as a function of \mMed alone for fixed \mDM is used to extract limits with various $M_{\chi}^{\prime}$, it is assumed that the $k$-factor at (\mMed, \mDM) is the same as the $k$-factor at (\mMed, $M_{\chi}^{\prime}$) within the model and scenario being studied. This assumption should be checked by the user at a few representative points using an NLO Monte Carlo simulation. If the $k$-factors are found to be inconsistent, a correction should be derived and applied or the bounds of the rescaling adjusted to remove the inconsistent points.

A second assumption of the rescaling method is that the effects of a signal on an analysis result are constant across the start and end points of a rescaling, less the effect of the cross section. Therefore a limit calculated for a signal point with a particular acceptance should not be rescaled to act as a limit on a signal point with very different acceptance, and a resonance search which is sensitive to the peak shape of a signal should reinterpret points only to signals with a similar peak shape and location. For \metplusx analyses, the acceptance is not overly sensitive to model parameters, but much like the $k$-factor assumption, the \MET for a few representative points towards the extremes of the rescaling should be checked to ensure the acceptance will not change meaningfully. For resonant final states, the relevant parameter is the mediator's intrinsic width. For intrinsic width to mass ratios significantly smaller than the analysis resolution, varying the couplings will have no effect on the mediator width so long as the values remain small, but when the analysis resolution is quite good or the signals become wider, the effects can be non-negligible. To account for this, the maximum intrinsic width to which a given limit is valid should be specified by the analyser, and the rescaling will not be applied for points outside this value. To support analyses like dilepton where the resolution is excellent, the rescaling code has been developed to support an input range of observed resonant limits applicable for different mediator widths, up to the maximum width limit provided. Details of this can be found in Section~\ref{subsec:dilepton}.

\section{Rescaling resonant final states}
\label{sec:resonant}

For resonant final states, as discussed in Ref.~\ref{sec:models}, the effects on the cross section of varying model parameters can be explored by considering the process $qq\rightarrow \text{med} \rightarrow f \bar{f}$ regardless of the presence of ISR. The kinematics, acceptance, and amount of background observed in the final state do depend on the presence of ISR, as well as strongly depending on the mass of the mediator. These factors enforce that any reinterpretation of limits must be within a single analysis final state, that is, must not alter the presence or absence of additional radiation, and cannot rescale a limit based on one mediator mass to constrain a different mediator mass. This last point is very clear when one considers how the exponentially decreasing background in a resonance search must result in a very different number of observed events corresponding to two different invariant masses. 

Resonant analysis limits can be reinterpreted across changing couplings, since varying couplings can affect only the cross section (the target for the rescaling) and the intrinsic width but not the invariant mass of an observed signal. Variations in intrinsic width will affect the detector signature of a signal point if that intrinsic width of the model is on the order of the experimental mass resolution. This is the case with high-resolution final states such as dilepton and can be handled by using analysis limits parameterised by intrinsic width. This will be discussed further in Section~\ref{subsec:dilepton}. 

One key point distinguishing visible from invisible final states is that resonant (visible) limits can also be reinterpreted across varying DM masses \mdm. With no coupling to DM present in the leading order interaction for these final states, changing \mdm will only affect the result through variation in the cross section and intrinsic width, as in the case of varying couplings. This means the analysis limits used as inputs to derive exclusions in the \mMed-\mdm plane can be one-dimensional and parameterised only as a function of \mMed.

The dijet and dilepton analyses typically publish a one-dimensional limit corresponding to a single model and set of couplings, which can take one of two standard forms. The first form is an observed 95\% confidence level upper limit on the cross section of new physics $\sigma_\text{obs}$ alongside a theoretical prediction for the cross section with any appropriate acceptance and efficiency factors $\sigma_\text{th}$, both as functions of \mMed, and whose intersection point marks the upper limit on \mMed for the model and scenario of the theory curve. The second form is a limit on $g_q$ or $g_l$ as a function of \mMed with all other couplings and \mdm fixed. These forms are in reality highly interconnected: in a dijet final state with all other couplings and \mdm fixed, for a specific \mMed, the the theoretical prediction for the cross section will scale proportionally to $g_q^2$ and the limit set on the coupling $g_{q,\text{lim}}$ is just that value for which the theoretical prediction matches the observed limit. The two limit forms can then be converted directly from one to the other via 
\begin{equation}
\label{eq:basic_gq}
g_{q,\text{lim}} = g_{q,\text{th}} \sqrt{\frac{\sigma_\text{obs}}{\sigma_\text{th}}}\,, 
\end{equation}
where $g_{q,\text{th}}$ is the value of $g_q$ in calculating $\sigma_\text{th}$. The equivalent for a dilepton final state, where the cross section scales proportionally to $g_qg_l$ and $g_q$ is fixed, is $g_{l,\text{lim}} = g_{l,\text{th}} (\sigma_\text{obs}/\sigma_\text{th})$.

Regardless of starting point, the goal is to produce an estimate of the \textit{exclusion depth} in a two-dimensional plane defined by \mMed and \mDM. We define exclusion depth $d_\text{ex}$ to be the ratio of the experimental cross section upper limit divided by the theoretical cross section for any point in the plane. The observed limit in the mass-mass plane is set by the contour $d_\text{ex}=1$. Points with $d_\text{ex} < 1$ are excluded while those with $d_\text{ex} > 1$ are not. With this information, rescaling to any model and scenario is as simple as multiplying the exclusion depth by the ratio of cross sections in the source and target scenario.

This requires a good approximation of the cross section for the processes in question. Fortunately, these resonant processes will be on-shell everywhere. For any \mMed of interest, in these visible final states, there will always be quarks or leptons with $m_f < 2\,\mMed$ into which the mediator can decay. Therefore this simple on-shell approximation can be applied for all points:
\begin{equation}
\label{eq:brrescaling}
\tilde{\sigma} \sim \frac{\Gamma_i\Gamma_f}{\Gamma_\text{tot}}\,,
\end{equation}
where $\Gamma_i$ and $\Gamma_f$ are the partial widths of the initial and final states respectively and $\Gamma_\text{tot}$ is the total intrinsic width of the mediator. This quantity is easy to compute since formulas for the partial widths are well-known in these models~\cite{ALBERT2019100377}.

The first step is to rescale the one-dimensional \mDM limit into a full grid of exclusion depths in (\mMed, \mdm).
In the initial one-dimensional limit on the cross sections, in which $d_\text{ex} = \sigma_\text{obs}/\sigma_\text{th}$, $\sigma_\text{th}$ corresponds to a specific set of couplings $g_q$, $g_\chi$, $g_l$ as well as a fixed \mdm and the \mMed value of each point on the x-axis. To get the exclusion depth in a target point with \mMedprime = \mMed but varied $g_q^\prime$, $g_\chi^{\prime}$, $g_l^\prime$ and \mdmprime, the original exclusion depth can simply be multiplied by the ratio of the cross section approximations for the two scenarios:
\begin{equation}
\label{eq:dijetrescaling_fromxsec}
d_\text{ex}^\prime = d_\text{ex} (\frac{\tilde{\sigma}}{\tilde{\sigma}^\prime}) = \frac{\sigma_\text{obs}}{\sigma_\text{th}} (\frac{\tilde{\sigma}}{\tilde{\sigma}^\prime})\,.
\end{equation}
Note that $\sigma_\text{th}$ and $\tilde{\sigma}$ are not the same: $\sigma_\text{th}$ is the full cross section including any analysis-relevant factors while $\tilde{\sigma}$ is an approximation missing various multiplicative factors and which is only relevant when compared to the same approximation at other model and scenario points.

When instead beginning from a limit on $g_q$ as a function of \mMed, the exclusion depth in the initial plot is defined by the relation in Eq.~\ref{eq:basic_gq}: $d_\text{ex} = (g_{q,\text{lim}}/g_{q,th})^2$. Combining this with Eq.~\ref{eq:dijetrescaling_fromxsec},
\begin{equation}
d_\text{ex}^\prime = (\frac{g_{q,\text{lim}}}{g_{q,th}})^2 \frac{\tilde{\sigma}}{\tilde{\sigma}^\prime}\,,
\end{equation}
where $g_{q,th}$ and $\tilde{\sigma}$ are defined for a theory curve which, combined with a matching observed limit curve, would result in the input $g_q$ limit. These are not defined as there is no input cross section curve, but their effect in fact cancels out: $\tilde{\sigma} \propto g_q^2$ for dijets and so selecting any arbitrary $g_q$ and evaluating $\tilde{\sigma}$ at the same will produce the same result. For convenience, we choose unity. Then using a $g_q$ limit at a hadron collider:
\begin{equation}
d_\text{ex}^\prime = g^2_{q,\text{lim}} (\frac{\tilde{\sigma}(g_q=1)}{\tilde{\sigma}^\prime})\,,
\end{equation}
all of which quantities can be calculated for a grid of (\mMed, \mdm) values from the one input curve. The equivalent exclusion depth extracted from a $g_l$ limit at a hadron collider with fixed $g_q$ is:
\begin{equation}
d_\text{ex}^\prime = g_{l,\text{lim}} (\frac{\tilde{\sigma}(g_l=1)}{\tilde{\sigma}^\prime})\,.
\end{equation}

Once one of the above formulas has been used to calculate $d_\text{ex}^\prime$ at each point of interest in the mass-mass plane for a single model and coupling scenario, the next step is to rescale to any other scenario or model of interest. This is achieved through multiplying the exclusion depth by a simple factor of $\tilde{\sigma}/\tilde{\sigma}^\prime$, where $\tilde{\sigma}$ is computed using Eq.~\ref{eq:brrescaling} for the set of couplings of the scenario from which to begin and $\tilde{\sigma}^\prime$ is in the target scenario.
Since the expressions for the partial widths include the factors that differ between vector and axial-vector couplings, these differences are included in the calculation and the simple $\tilde{\sigma}/\tilde{\sigma}^\prime$ rescaling can be used to convert between the two models. The dijet and dilepton resonant final states and their variants do not provide strong constraints on scalar and pseudoscalar models, where other final states are dominant. Conversion between vector and axial-vector as well as across couplings within each model therefore covers the full space of interesting rescaling possibilities for these two models.

The rescaling method discussed here for resonant final states is based on a method used by CMS collaborators to present the dijet and dilepton limits in the CMS summary plots since the beginning of Run 2.
The technique has been since been adopted by ATLAS for the presentation of their limit summaries. This paper makes one small extension on the well-established rescaling technique by enabling limits to be extracted from $g_q$ limits as a function of mass regardless of whether \mdm is taken to be decoupled or light. 

\subsection{Dijet and dijet+X}
\label{subsec:dijet}

The simplest example of resonant limit rescaling is for an analysis when the intrinsic mediator width-to-mass ratio is smaller than or comparable to the experimental mass resolution. This is common for dijet analyses where the mass resolution is on the order of 4\% at $\mMed = 1$ TeV and 3\% at $\mMed = 2$ TeV~\cite{dijetISR,dijet139_atlas}. The rescaling methods derived in Section~\ref{sec:resonant} are illustrated using the results of the CMS dijet analysis in 36 \ifb of data~\cite{CMS:2018mgb}. Since CMS has been using these resonance search rescaling methods for some time, the mass-mass exclusion limits presented in that paper are already derived using the procedures given here and beginning from the limit on $g_q$ as a function of \mMed with no coupling to dark matter. To validate the equivalency of rescaling from cross-section or $g_q$ limits with or without decoupled dark matter, Figure~\ref{fig:br_scaling} shows that the same mass-mass limits can be obtained beginning from the cross-section limit and theory curve corresponding to $g_q = 0.25$, $g_\chi=1.0$, \mdm$= 1$ GeV instead. The observed limit on $\sigma \times B \times A$ is taken from the public data while the theory curve and is visually extracted from the limit plot.

\begin{figure}[htp!]
  \begin{center}
  \begin{subfigure}[b]{0.49\textwidth}  
    \includegraphics[width=\textwidth]{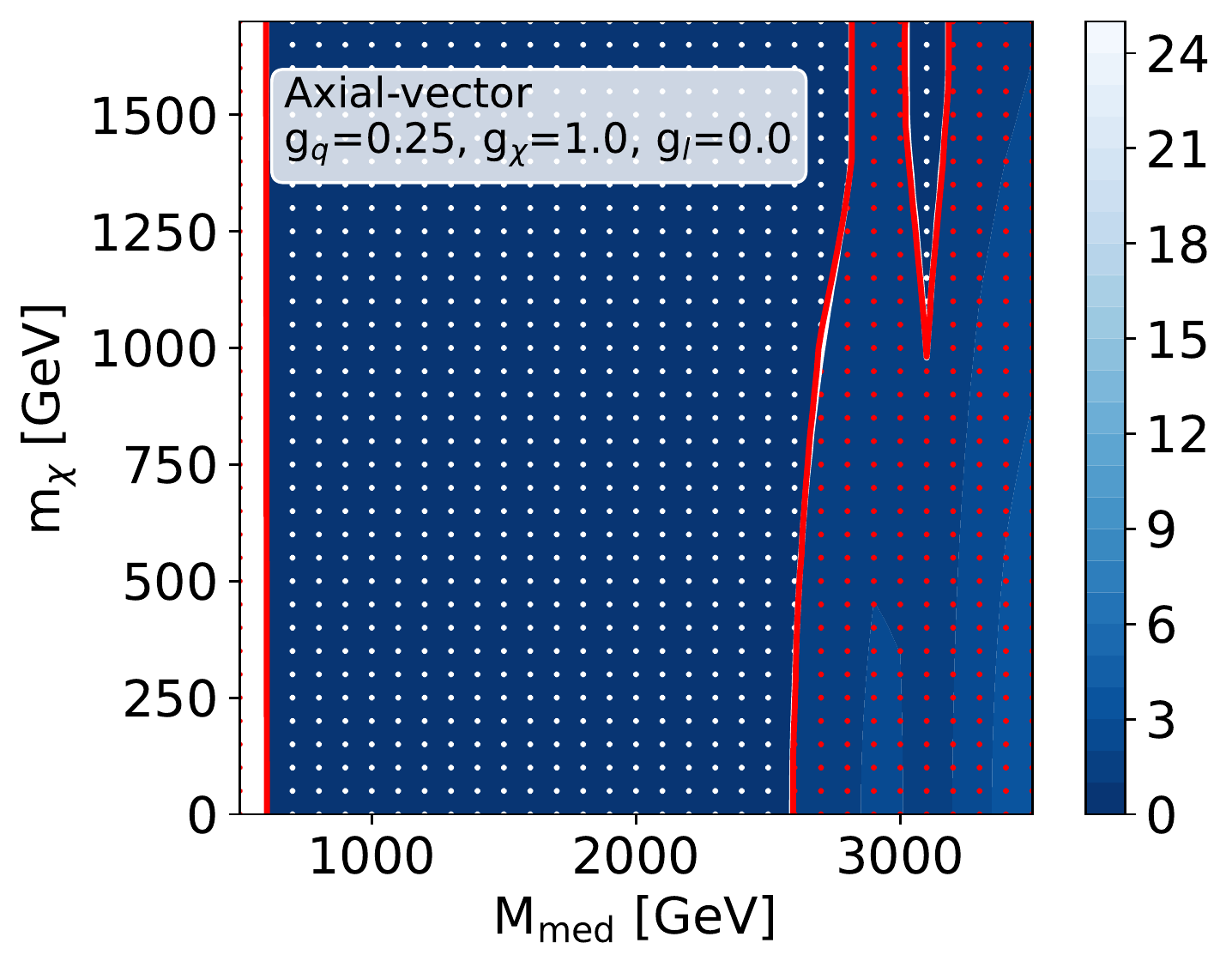}
    \caption{}
    \label{subfig:dijet_A1}
  \end{subfigure}
  \begin{subfigure}[b]{0.49\textwidth}  
    \includegraphics[width=\textwidth]{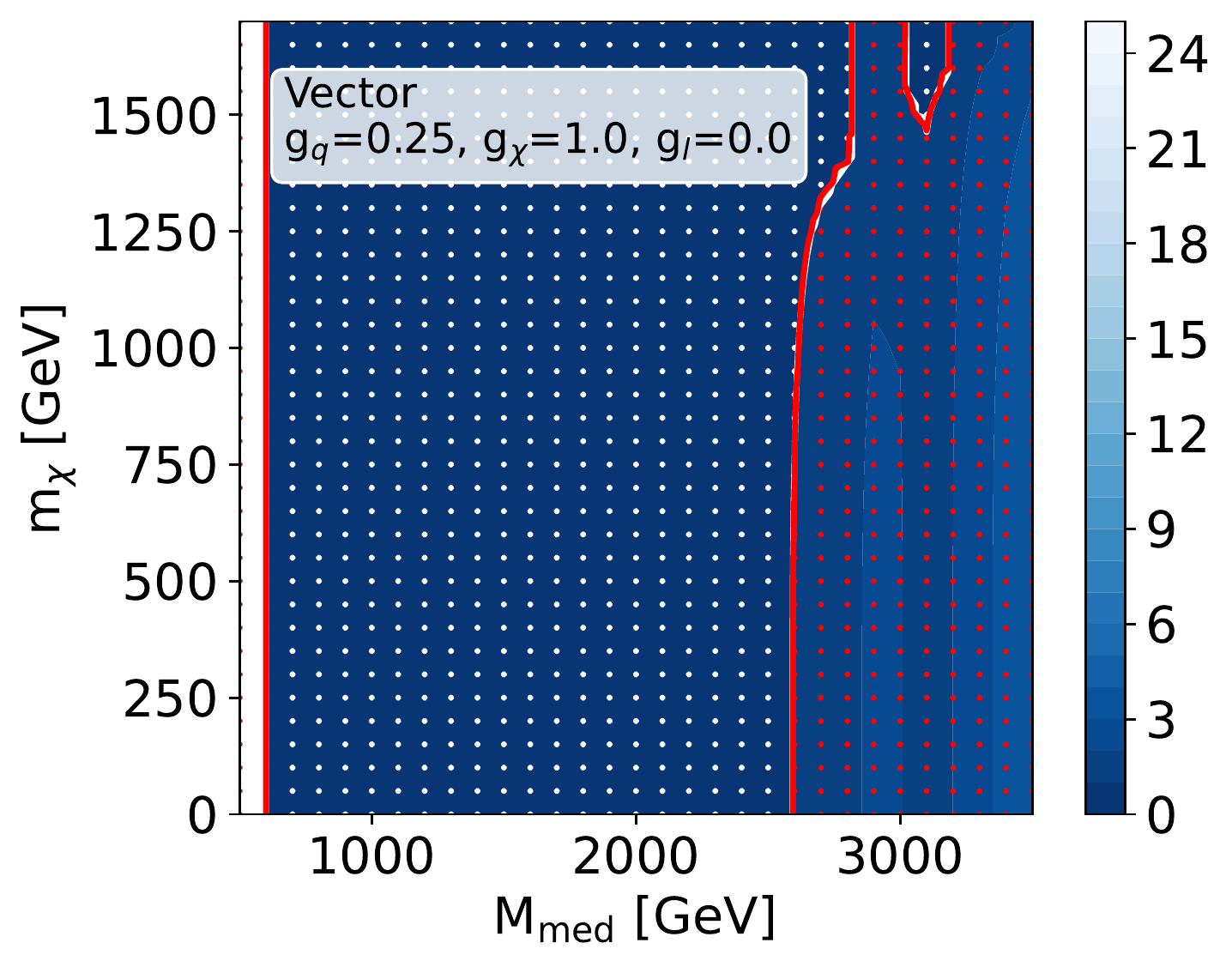}
    \caption{}
    \label{subfig:dijet_V1}  
  \end{subfigure}
  \caption{Exclusion contours in the (\mMed, \mdm) plane for axial-vector mediators (\subref{subfig:dijet_A1}) and vector mediators (\subref{subfig:dijet_V1}) with couplings of $g_q=0.25, g_\chi=1.0, g_l = 0.0$ using the CMS dijet analysis results. The shape of this exclusion surface is shown as a blue gradient. Each point at which the exclusion depth was calculated is plotted as a dot: excluded points are plotted in white, while points which are not excluded are plotted in red. The solid red curves correspond to the published CMS results in these planes computed using the rescaling methods discussed here and beginning from the limit on $g_q$ as a function of mass for a mediator with no coupling to dark matter. The solid white curves correspond to the result derived using the rescaling methods discussed here and beginning from the observed limit on $\sigma \times B \times A$ as a function of \mMed and the corresponding theoretical prediction for a mediator with coupling $g_\chi=1$ and a dark matter particle with mass $\mdm = 1$ GeV. The level of agreement between the two contours is a measure of the similarity in performance between the rescaling procedures beginning from these two different initial limits.}
  \label{fig:br_scaling}
  \end{center}
\end{figure}

Determining the largest intrinsic width for which the input observed limit is applicable is the responsibility of each user of the provided software. The software can then take this largest acceptable width as an input and will only return rescaled limits when the point being tested falls within this region of validity. This is intended to avoid accidental reporting of a limit in an unjustified context. For cases where a single observed limit curve cannot apply to all the cases of interest, a modification of this approach is applied, as described in the following section.

\subsection{Dilepton and other high-resolution resonances}
\label{subsec:dilepton}

In many resonance search situations, the experimental mass resolution is very good and can be significantly smaller than the intrinsic width of these mediators, leading to observed limits that would vary significantly for different mediators with the same \mMed. One example is a di-electron final state, in which the mass resolution is about 1\% for all mediator masses over $\sim800$ GeV~\cite{dilepton_139}.
In this case, multiple observed limits can be provided corresponding to different intrinsic mediator widths. During the rescaling process, at each point to be studied, the observed limit used to compute $d_\text{ex}$ will be found by interpolating between the input observed limits at the two nearest widths to that of the studied point. The implementation provided with this paper uses a linear interpolation between each neighbouring pair of observed limit inputs. All points with intrinsic widths smaller than that of the narrowest provided limit curve will take their observed values from that narrowest limit curve with no interpolation or extrapolation. Points with a larger intrinsic width than that of the widest provided limit curve will be considered to fall outside the bounds of analysis applicability and will not return rescaled limits. 

This method can be applied to dijet final states as well, or to any resonant final state where it is desirable to eliminate the assumption that the observable signal shape and acceptance are unchanged as a function of DM mass and coupling values. By providing a range of limits corresponding to various intrinsic width-to-mass ratios for a single mediator mass point, the most appropriate one can be chosen for each coupling and DM mass.

\section{Rescaling \metplusx final states}
\label{sec:monox}

For \metplusx signatures, the relevant final state involves decay of a mediator to dark matter particles, and therefore the off-shell case has to be handled appropriately.
The approximations used for visible resonant final states in Section~\ref{sec:resonant} require a well-defined decay width to the final state $\Gamma_f$, which goes to zero at $\mMed=2\mDM$ for an invisible final state. The approximation in fact loses validity before this diagonal is fully reached, as the transition across the on-shell to off-shell boundary is smooth. A method for re-scaling \metplusx signatures has therefore been developed with the goal of handling this transition smoothly such that it is applicable in all regimes.

Previously, ATLAS and CMS \metplusx analyses have generated a grid of signal points in one of the four benchmark scenarios at full reconstruction level and used these to determine the region of (\mMed, \mDM) space excluded in that scenario~\cite{atlas_monojet_36ifb, cms_monojet_12ifb}. Three additional grids of signal points are then generated at LO and particle level in order to obtain cross section estimates for each point in the additional three benchmark scenarios. The ratio of cross sections between the points in the original scenario and the target scenarios is used to scale the limits, resulting in a new estimate of the excluded and non-excluded points within each target scenario. 
The goal of these studies is to use the same starting information, a full grid of results in the (\mMed, \mDM) plane for one coupling scenario, but rescale to another target scenario quickly and mathematically instead of through signal generation.
Note that unlike in the resonant case where a one-dimensional limit in \mMed is sufficient, a two-dimensional grid is required here. Because dark matter particles appear in the final state for \metplusx signatures, both \mdm and \mMed affect the \MET distribution and analysis acceptance, and it is therefore not possible to rescale a limit across either of these dimensions.

The semi-analytical rescaling method developed here for \metplusx signatures has two separate components. One can be used to rescale an initial scenario to another set of couplings within the same overall model, while the other should be used when a scenario in one model is rescaled to a target in a different model (e.g. vector mediator to axial-vector mediator).

In the ATLAS/CMS Dark Matter Forum whitepaper~\cite{ABERCROMBIE2020100371} defining the simplified models studied here, it is specified that the cross section scaling can be estimated using the integral of the Breit-Wigner propagator for the mediator. Several approximations of this integral are given, corresponding to the different regimes of on-shell mediators, off-shell mediators, and effective field theories. In order to smoothly handle the on-shell to off-shell transition, we instead use the full integral of the propagator over the permitted phase space $s \geq 4 \mDM^2$:
\begin{align}
\mathcal{I}_{\text{prop}} =&\  g_q^2 g_\chi^2 \int_{4\mDM^2}^{\infty} \frac{ds}{(s-\mMed^2)^2+\mMed^2\Gamma^2} \\
=&\  \frac{g_q^2 g_\chi^2}{\Gamma\mMed} \left(\frac{\pi}{2} + \arctan{\left(\frac{\mMed^2 - 4 \mDM^2}{\Gamma\mMed}\right)} \right)\,.
\end{align}
Coupling dependence comes not only from the explicit $g_q^2 g_\chi^2$ factor but is also embedded in the value of $\Gamma$ (see width definitions in Ref.~\cite{ALBERT2019100377}).
$\mathcal{I}_{\text{prop}}$ is easily analytically calculable and provides a robust estimate of the cross section when comparing two different coupling scenarios, so long as both assume the same mediator type. A starting grid of exclusion depths can then be easily rescaled to a target scenario of the same model by multiplying each $d_\text{ex}$ value by the ratio of $\mathcal{I}_{\text{prop}}$ in the initial and target scenarios at that point.

The $\mathcal{I}_{\text{prop}}$-based rescaling procedure is illustrated in Figure~\ref{fig:propagator_scaling} using the observed limits from the ATLAS $\met+\gamma$ analysis in 36 \ifb of data~\cite{monophoton}. Circles show the locations of the signal points used. Each signal point has a $z$-axis value of theoretical cross section divided by excluded cross section; white points are those with $z<1$ and are excluded while red points have $z>1$ and are not excluded. The red curves show the exclusion contours reported by the published analysis in each plane. The blue shades are a linear interpolation between the points used to illustrate the cross section exclusion surface. In Figure~\ref{subfig:monophoton_A1} no reinterpretation is done: the values at each point are taken directly from the paper and overlaid with the corresponding exclusion curve. In Figure~\ref{subfig:monophoton_A2} the values at each point are calculated using the $\mathcal{I}_{\text{prop}}$ scaling procedure starting from the values in Figure~\ref{subfig:monophoton_A1}. The white and red colour coding of points is based on these rescaled values, as is the linear interpolation in blue. The good agreement between the contour curve taken from the paper and the points marked as excluded by the rescaling method serves as a validation of the rescaling.

\begin{figure}[htp!]
  \begin{center}
  \begin{subfigure}[b]{0.49\textwidth}  
    \includegraphics[width=\textwidth]{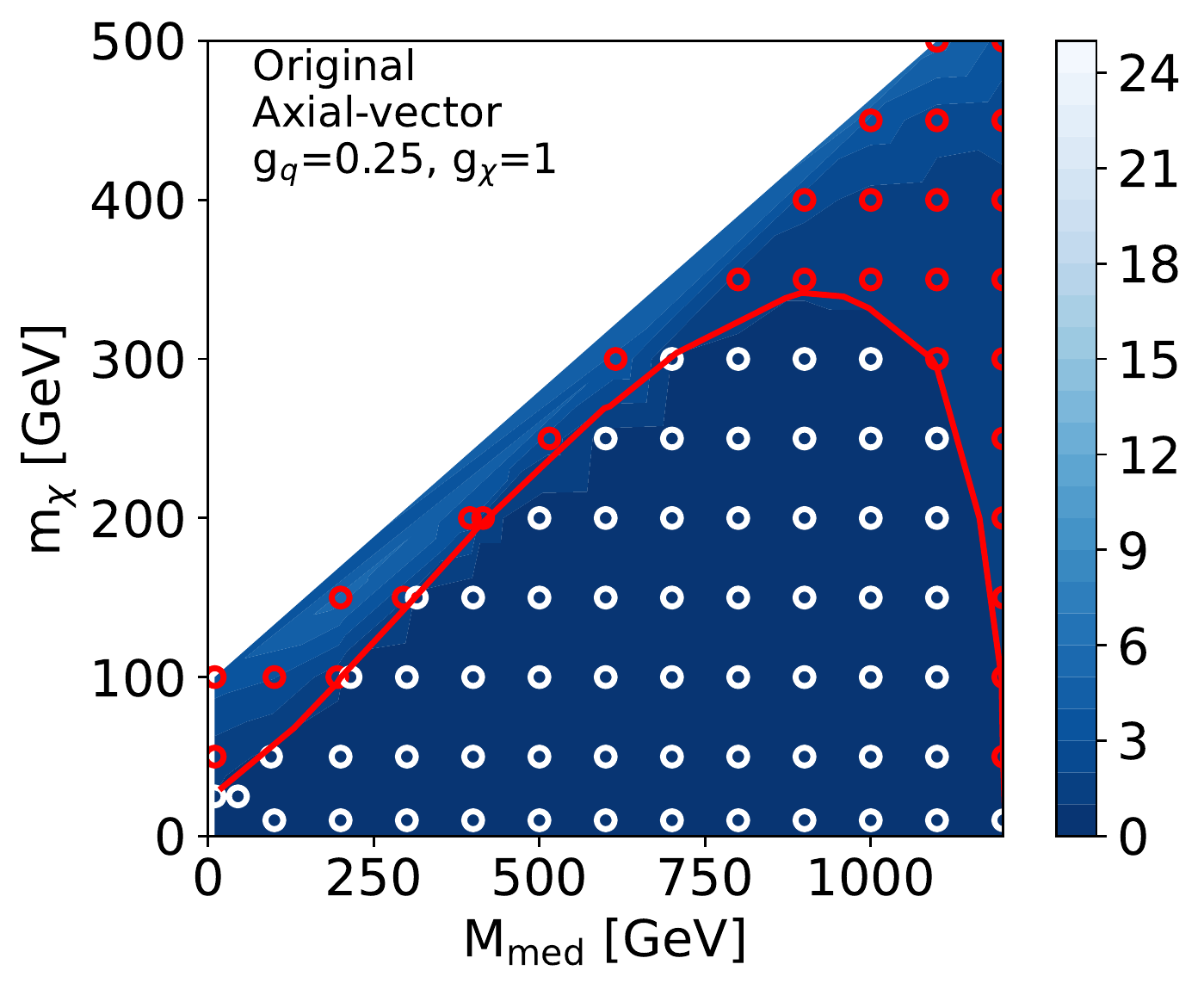}
    \caption{}
    \label{subfig:monophoton_A1}
  \end{subfigure}
  \begin{subfigure}[b]{0.49\textwidth}  
    \includegraphics[width=\textwidth]{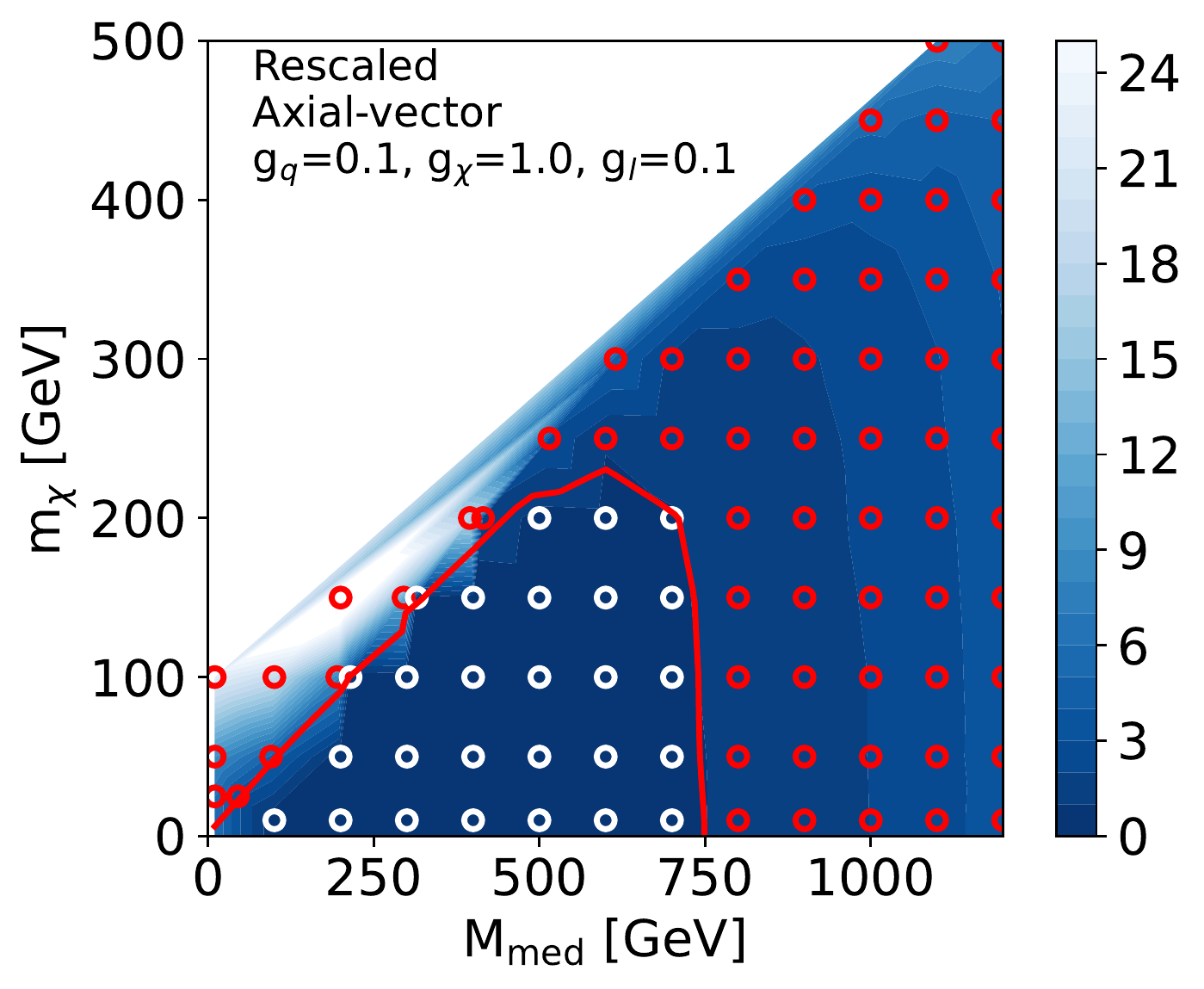}
    \caption{}
    \label{subfig:monophoton_A2}  
  \end{subfigure}
  
  \caption{Original cross section limits with couplings $g_q=0.25,~g_\chi=1.0,~g_l = 0.0$ (\subref{subfig:monophoton_A1}) and rescaled cross section limits with couplings $g_q=0.1, g_\chi=1.0, g_l = 0.1$ (\subref{subfig:monophoton_A2}) for an axial-vector mediator model using the ATLAS $\met+\gamma$ analysis results~\cite{monophoton}. Limits are presented as a function of mediator mass and dark matter mass for fixed coupling values. At each signal point, the value of $d_\text{ex} = \sigma_{\text{theory}}/\sigma_{\text{excluded}}$ is calculated: points where $d_\text{ex}<1$ are excluded and plotted in white, while points where $d_\text{ex}>1$ are not excluded and plotted in red. A linear interpolation between the $d_\text{ex}$ values at the points illustrates the shape of this exclusion surface and is shown as a blue gradient. The solid red curves are taken from the published $\met+\gamma$ results and show the exclusion contour $d_\text{ex} = 1$. The level of agreement between the excluded points and the published contour in~(\subref{subfig:monophoton_A2}) is a measure of the similarity in performance between the $\mathcal{I}_{\text{prop}}$ scaling procedure and the Monte Carlo based method used in the ATLAS analysis.}
  \label{fig:propagator_scaling}
  \end{center}
\end{figure}

Rescaling cross sections using a ratio based on $\mathcal{I}_{\text{prop}}$ is no longer sufficient when the target scenario uses a different mediator model than the initial scenario. Additional mass-dependent terms in the cross section vary between models, so while they cancel within a single model and can be ignored in favour of the simple propagator-based expression, they must be correctly accounted for when rescaling from one model to another. We therefore calculate a more complete cross section whose ratio can be used to perform cross-model rescaling.

As discussed in Section~\ref{sec:models}, the LO signal contribution to each \metplusx analysis can be calculated from the $qq\rightarrow \text{med} \rightarrow \chi \bar{\chi}$ process independent of additional ISR radiation. Therefore, to calculate the LO scale factor translating between two scenarios with different mediator models, the cross section for that single diagram in both models is sufficient. The parton-level cross sections as functions of the interaction scale $S$ are given by the following equations, discounting scale factors which will cancel in the ratio:

\begin{equation}
\sigma_V(S) \propto \frac{g_q^2 g_\chi^2 (S + 2\mDM^2)\sqrt{S-4\mDM^2}}{(\Gamma^2\mMed^2 + (\mMed^2 - S)^2)\sqrt{S}}
\end{equation}
and
\begin{equation}
\sigma_{AV}(S) \propto \frac{g_q^2 g_\chi^2 (S-4\mDM^2)^{3/2}}{(\Gamma^2\mMed^2 + (\mMed^2 - S)^2)\sqrt{S}}\,,
\end{equation}
where $\Gamma$ is the total width of the mediator (non-zero even when decays to DM are off-shell).

To calculate cross sections with sufficient accuracy for the rescaling procedure, a hadron-level quantity must be determined accounting for parton distribution functions (PDFs). The full cross section is thus calculated semi-analytically by performing a two dimensional integral over the longitudinal momentum fractions $x_1, x_2$ of the interacting partons within the allowed range. Where $S$ is the squared centre-of-mass energy (i.e. $\sqrt{S} = 13$ TeV), $\hat{s} = x_1 x_2 S$ is the scale of the interaction, and $f^a$ and $f^b$ are the PDFs of the incoming partons, the total cross section is
\begin{equation}
\sigma_{V/AV}^{\text{tot}} = \int_{4 \mDM^2}^{\text{S}} \int_{4 \mDM^2}^{x_2 S}  f^a(x_1,\hat{s})  f^b(x_2,\hat{s}) \sigma_{V/AV}(\hat{s}) dx_1 dx_2\,.
\end{equation}
This integral is performed numerically in the two models and the ratio at each point is taken as the scale factor to convert between them. The implementation uses LHAPDF and so provides a wide range of PDF sets and flavour schemes, although results are found to be fairly independent of the selection~\cite{Buckley:2014ana}.

A demonstration of the $\sigma_{V/AV}^{\text{tot}}$-based scaling procedure is given in Figure~\ref{fig:pdf_scaling}. The $z$-value at each point in Figure~\ref{subfig:monophoton_V2} is found using the ratio of the vector and axial-vector cross sections $\sigma_{V}^{\text{tot}}$ and $\sigma_{AV}^{\text{tot}}$ applied to the axial-vector limits from Figure~\ref{subfig:monophoton_A1}. The success of this rescaling method is again illustrated by the agreement between the excluded points (white) determined by the rescaling and the solid curve obtained from the analysis paper. Figure~\ref{subfig:monophoton_V2} is created by further applying a $\mathcal{I}_{\text{prop}}$ rescaling to the results in Figure~\ref{subfig:monophoton_V1}. Even after this iterative application of $\sigma_{V/AV}^{\text{tot}}$ and $\mathcal{I}_{\text{prop}}$ rescaling, the agreement between the points and the published results is fair.

\begin{figure}[htp!]
  \begin{center}
  \begin{subfigure}[b]{0.49\textwidth}  
    \includegraphics[width=\textwidth]{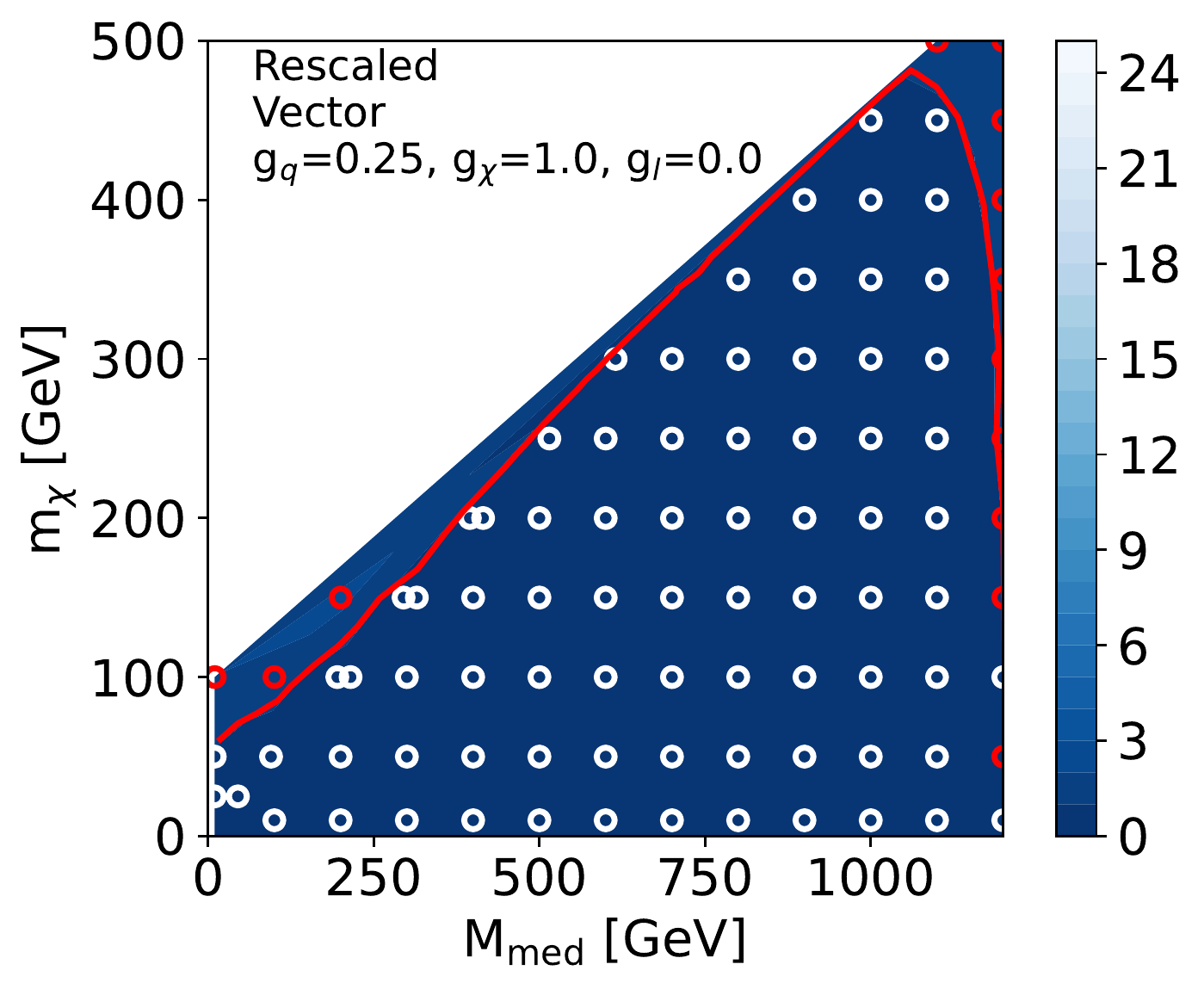}
    \caption{}
    \label{subfig:monophoton_V1}
  \end{subfigure}
  \begin{subfigure}[b]{0.49\textwidth}  
    \includegraphics[width=\textwidth]{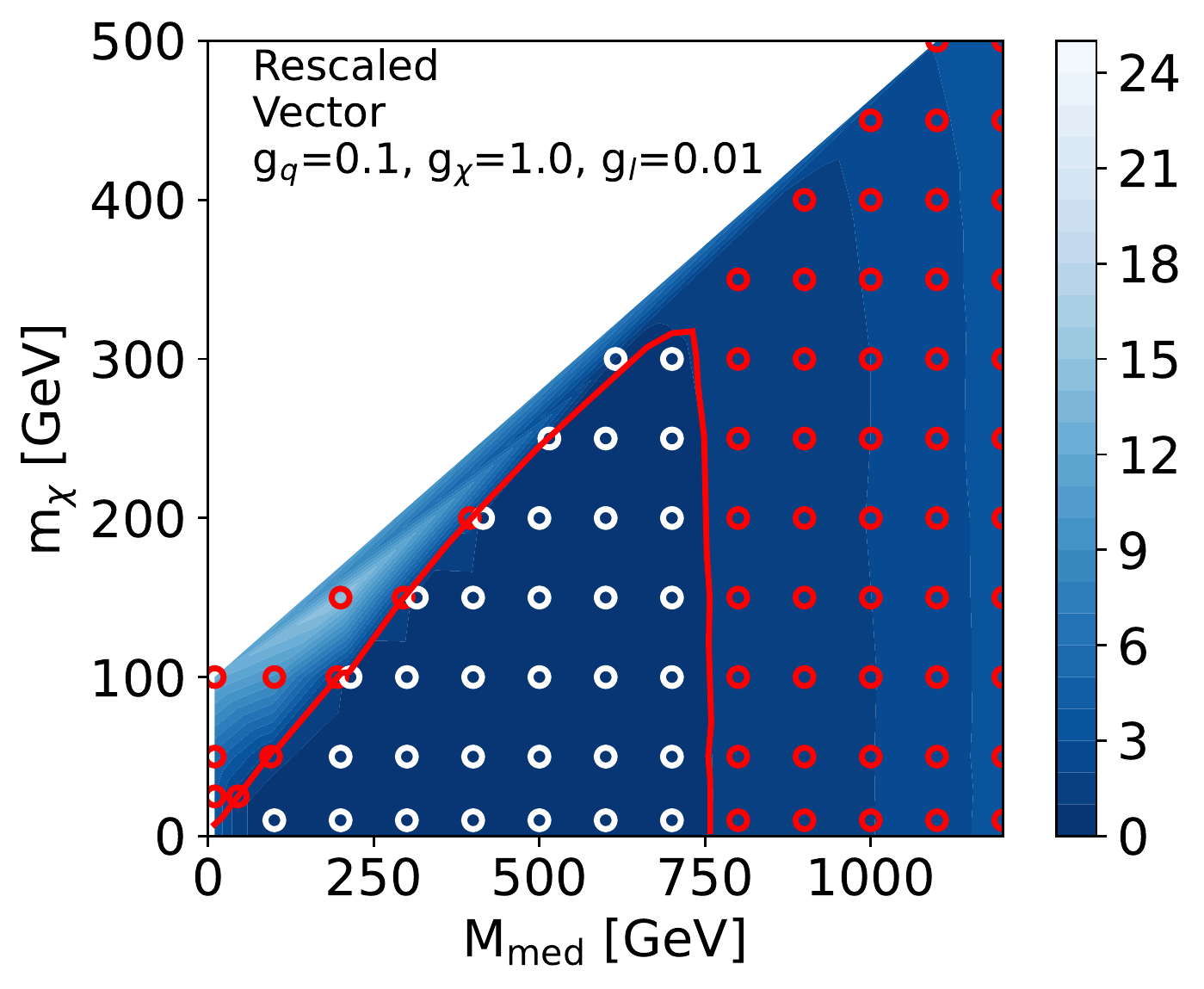}
    \caption{}
    \label{subfig:monophoton_V2}  
  \end{subfigure}
  
  \caption{Rescaled cross section limits with couplings $g_q=0.25, g_\chi=1.0, g_l = 0.0$ (\subref{subfig:monophoton_V1}) and $g_q=0.1, g_\chi=1.0, g_l = 0.01$ (\subref{subfig:monophoton_V2}) for a vector mediator model using the ATLAS $\met+\gamma$ analysis results. Limits are presented as a function of mediator mass and dark matter mass for fixed coupling values. At each signal point, the value of $d_\text{ex} = \sigma_{\text{theory}}/\sigma_{\text{excluded}}$ is calculated: points where $d_\text{ex}<1$ are excluded and plotted in white, while points where $d_\text{ex}>1$ are not excluded and plotted in red. A linear interpolation between the $d_\text{ex}$ values at the points illustrates the shape of this exclusion surface and is shown as a blue gradient. The solid red curves are taken from the published $\met+\gamma$ results and show the exclusion contour $d_\text{ex} = 1$. The level of agreement between the excluded points and the published contour is a measure of the performance of the $\sigma_{V/AV}^{\text{tot}}$ scaling procedure in~(\subref{subfig:monophoton_V1}) and of the combination of both procedures in~(\subref{subfig:monophoton_V2}).
  }
  \label{fig:pdf_scaling}
  \end{center}
\end{figure}

The full recommended procedure for rescaling \metplusx analysis results is therefore to use the $\sigma_{V/AV}^{\text{tot}}$-based rescaling procedure to convert from the original model with one type of mediator into a single scenario with the desired new mediator type (e.g. vector mediator to axial-vector mediator). The $\mathcal{I}_{\text{prop}}$-based rescaling procedure can then be used to convert to other coupling scenarios with the same mediator type.

\section{Combined examples of coupling-scaled exclusion plots}

An example of the opportunities which rescaling across couplings presents is shown in Figure~\ref{fig:combined}. The limits from the CMS dijet and ATLAS monophoton analyses studied above are combined into single exclusion contours for a fixed set of couplings and a range of couplings are overlaid. For this figure, $g_\chi$ and $g_l$ are held constant while $g_q$ is gradually reduced. Both dijet and monophoton exclusion limits become weaker as $g_q$ decreases. Such a detailed study of the effect of couplings on exclusion limits is possible because of the drastically reduced computing time involved in determining each rescaling.

\begin{figure}[htp!]
  \begin{center}
  \begin{subfigure}[b]{0.49\textwidth}  
    \includegraphics[width=\textwidth]{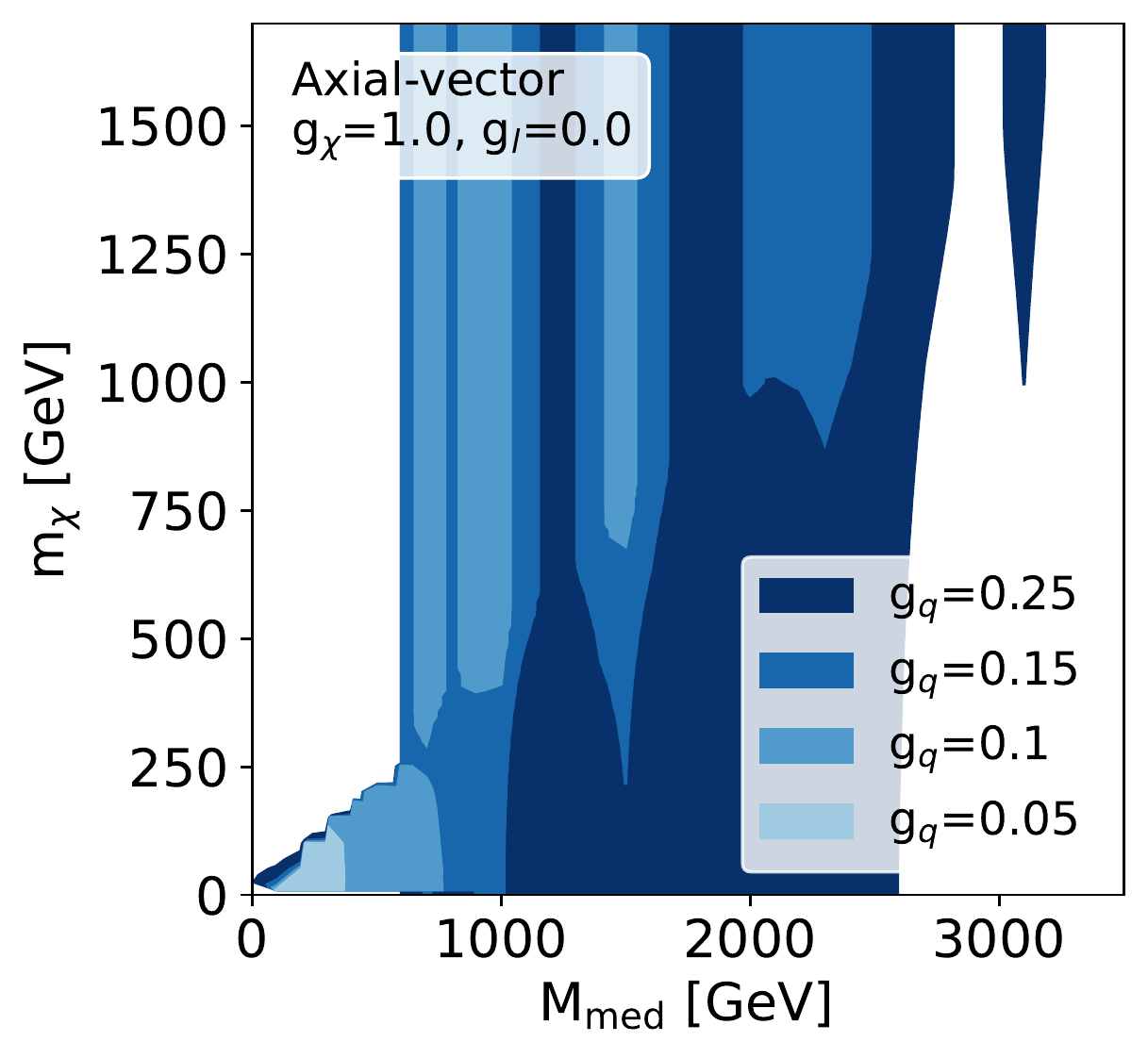}
    \caption{}
    \label{subfig:combined_a}
  \end{subfigure}
  \begin{subfigure}[b]{0.49\textwidth}  
    \includegraphics[width=\textwidth]{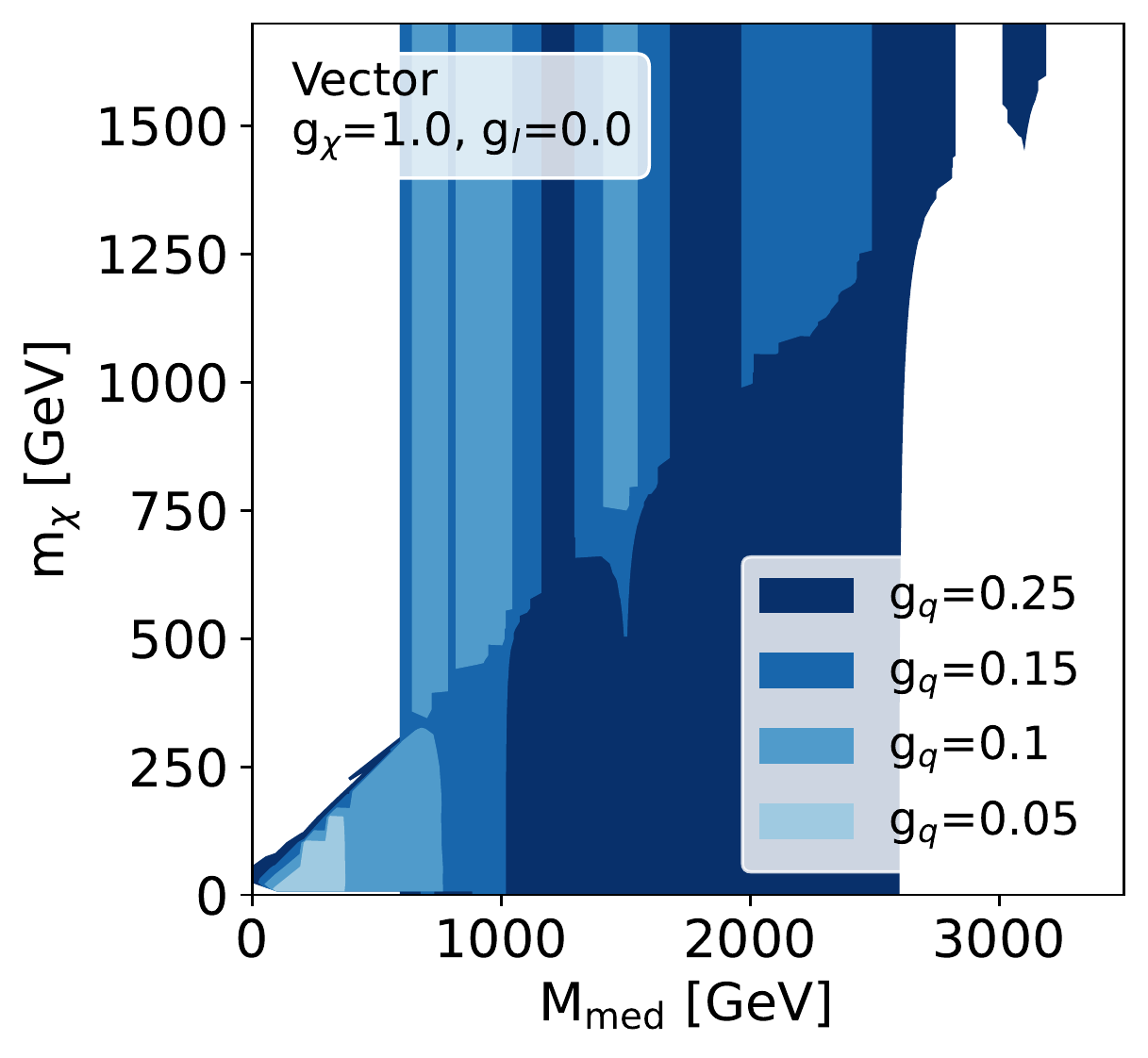}
    \caption{}
    \label{subfig:combined_v}  
  \end{subfigure}
  \caption{Exclusion limits of the dijet and monophoton analyses combined, corresponding to 36 fb$^{-1}$ of LHC data, when the coupling to quarks is varied. The dark matter coupling is held fixed at $g_\chi=1.0$ and there is no coupling between the mediators and leptons. The shaded region is excluded at the corresponding $g_q$ value. Axial-vector (\subref{subfig:combined_a}) and vector (\subref{subfig:combined_v}) mediators show similar patterns, though with a more pronounced cutoff along the $\mMed = 2 \mdm$ line for axial-vector mediators. All exclusion regions shrink as $g_q$ decreases.
  }
  \label{fig:combined}
  \end{center}
\end{figure}

Similar plots can be made varying $g_\chi$ or $g_l$ instead. Decreasing $g_l$ greatly reduces the power of dilepton exclusions while slightly increasing the power of dijet exclusions, since the dijet signatures then takes a higher fraction of the branching ratio; it has a smaller but nonzero impact on \metplusx exclusion curves. Decreasing $g_\chi$ increases the power of visible final state exclusions and decreases that of \metplusx analyses. 

\section{Relic densities in DM simplified models}

Calculation of the dark matter relic density involves computing the cross section for dark matter annihilation to standard model particles. In this scenario, two production modes tend to dominate, single mediator production and double mediator production. Single mediator production occurs when the dark matter annihilates with itself to create a single mediator that then decays to standard model particles. This production mode is the inverse of the production mode probed when searching for \metplusx final states, and is resonantly enhanced along the $2 \mDM=\mMed$ line. The second production mode arises from double mediator production. Diagrams for both of these production modes are shown in Figure~\ref{fig:relicprod}~\cite{Albert:2017onk}.

\begin{center}
\begin{figure}[!t]
\centering
\includegraphics[width=0.78\textwidth]{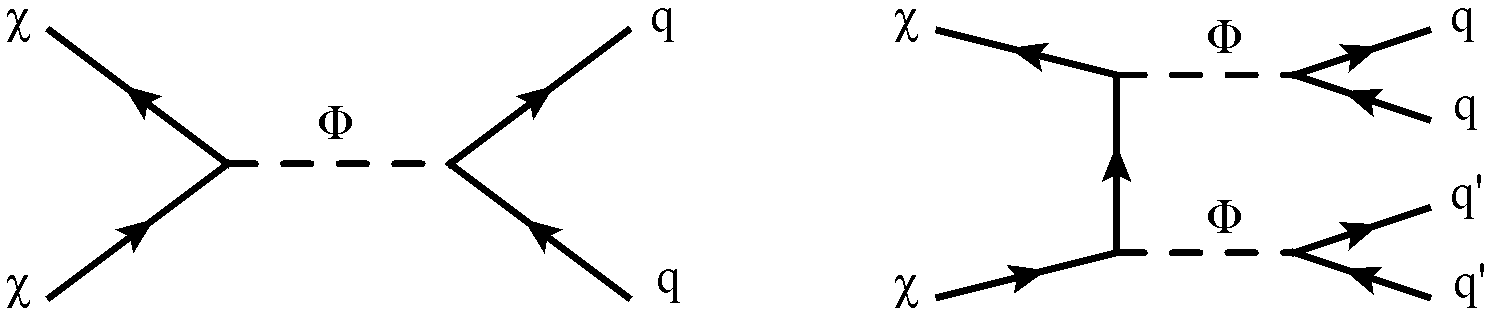} 
\vspace{4mm}
\caption{Feynman diagrams of DM $s$-channel annihilation to quarks (left) and $t$-channel annihilation to a pair of mediators subsequently decaying to quarks (right). The exchanged~$\Phi$ particle(s) can be either (pseudo-)scalar or (axial-)vector mediator(s). 
}
\label{fig:relicprod}
\end{figure}
\end{center}
  
When computing the relic density, we can approximate broad features of the calculation by understanding the form of the leading order production cross section. The leading order cross section can be written as
\begin{equation}
\label{eq33}
  \sigma_{{\rm ann},s}^V \cdot v = \sum_{q} \frac{N_c^q \hspace{0.125mm} g_{\chi}^2 \hspace{0.25mm} g_q^2 \hspace{0.5mm} \beta_q}{2 \pi} \,  \frac{2 \mdm^2+ m_q^2}{\left (\mMed^2 - 4 \mdm^2 \right )^2 + \mMed^2 \Gamma_{\rm med}^2} \,, 
\end{equation}
for the single mediator vector production, and it can be written as

\begin{equation}
\label{eq37} 
  \sigma_{{\rm ann},t}^V \cdot v  =  \frac{ g_{\chi}^4   \hspace{0.5mm} \beta_{\rm med}}{4 \pi}  \, \frac{\mdm^2 - \mMed^2 }{\left (\mMed^2 - 2 \mdm^2 \right )^2}\,,
\end{equation}
for the double mediator production.

When we consider single mediator production, if we take $ \sigma_{{\rm ann},t}^V \cdot v $ to be proportional to the thermalized cross section obtained from the full relic density calculation, and we fix both dark matter mass \mdm and mediator mass \mMed, we find that both equations above can be written in the form of a quadratic equation. Namely, for some constants $A,B,C,D,E,F$, we can write
\begin{align}
A g_{q}^2g_{\chi}^2 + B g_{q}^2+Cg_{\chi}^2+D g_{q}^4+ E g_{\chi}^4+F=0
\end{align}
This also follows from the fact that the widths for SM particles $\Gamma_{\rm SM}\propto g_{q}^2$ while the width for dark matter  $\Gamma_{\chi}\propto g_{\chi}^2$. From this form, we observe that if $g_{\chi}$ is a constant, then this equation immediately reduces to a quadratic equation in $g_{q}^2$, and likewise for $g_{\chi}^2$ when $g_{q}$ is a constant. Furthermore, the quartic terms $D$ and $E$ are also positive. As a consequence, when all couplings are fixed with the exception of either $g_{q}$ or $g_{\chi}$ there is a well defined coupling minimum. For couplings smaller than the minimum the relic is larger, and the same is true for couplings larger than the maximum. By scanning the couplings, we can find both the minimum and maximum allowed coupling that would satisfy the dark matter relic density.

Figure~\ref{fig:mincoupling} shows the result of a procedure whereby we scan the standard model coupling, compute the relic density, and find the minimum allowed coupling that would not overproduce dark matter. We present this plot for each of the simplified models used within the LHC Dark Matter Working Group. We find that for large regions of phase space, in particular light dark matter with a heavy mediator, there is no solution that will produce an underabundance of dark matter. Additionally, we find that couplings of order $g_{q}=1$ are required to satisfy the dark matter relic density for spin-0 scalar and pseudo-scalar mediators, and couplings of order $g_{q}=0.1$ or larger are needed to explain spin-1 vector and axial-vector mediators.

The bounds on these models are clearly substantial, and help to serve as a guide. We would like to stress that as models become more complicated these bounds can change substantially. However, this provides a clear set of benchmarks that can viewed as a target for both invisible, for light dark matter, and visible, for heavy dark matter, searches. 

\begin{figure}[htp!]
  \begin{center}
  \begin{subfigure}[b]{0.49\textwidth}  
    \includegraphics[width=\textwidth]{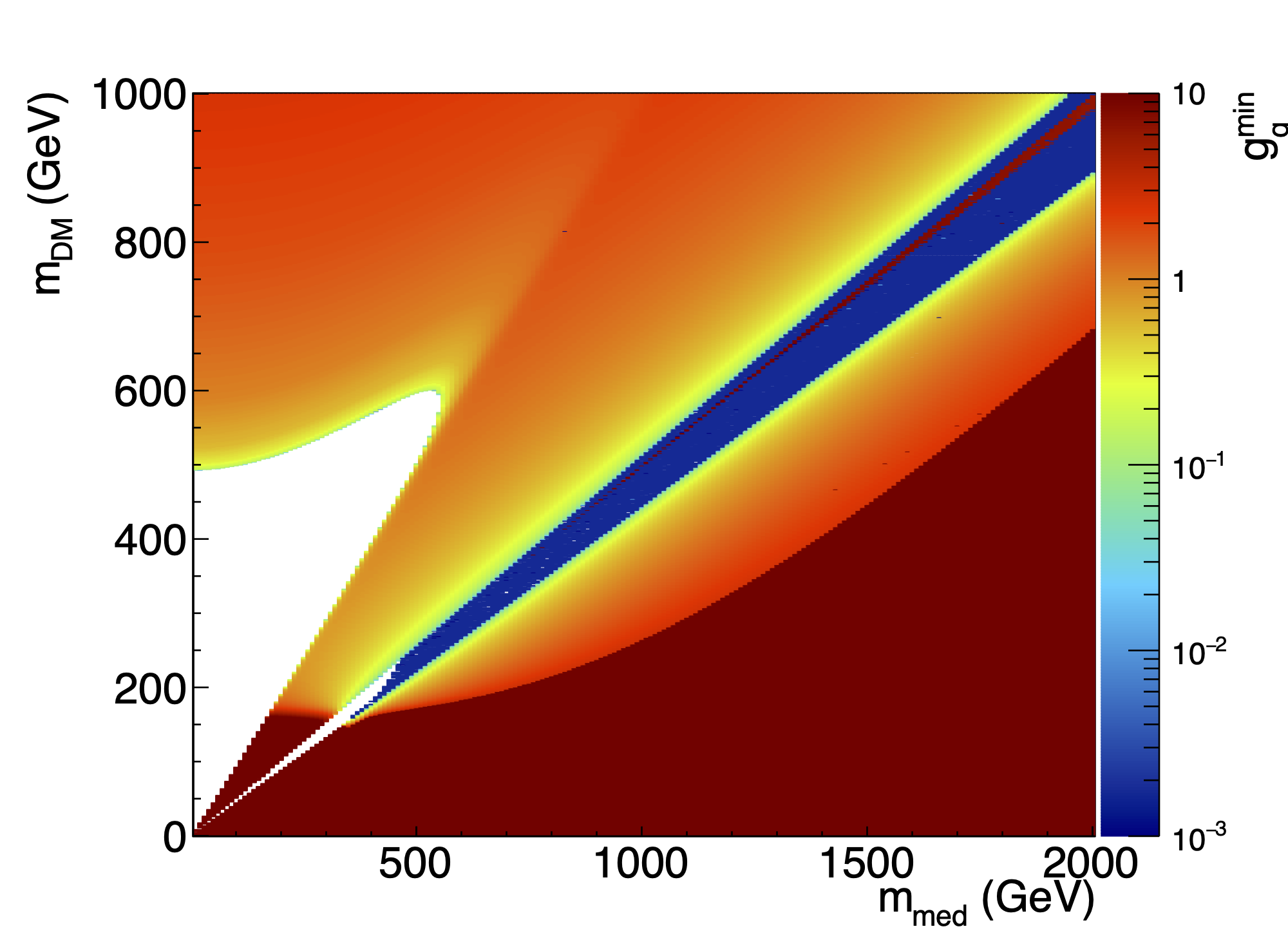}
    \caption{}
    \label{subfig:scalar}
  \end{subfigure}
  \begin{subfigure}[b]{0.49\textwidth}  
    \includegraphics[width=\textwidth]{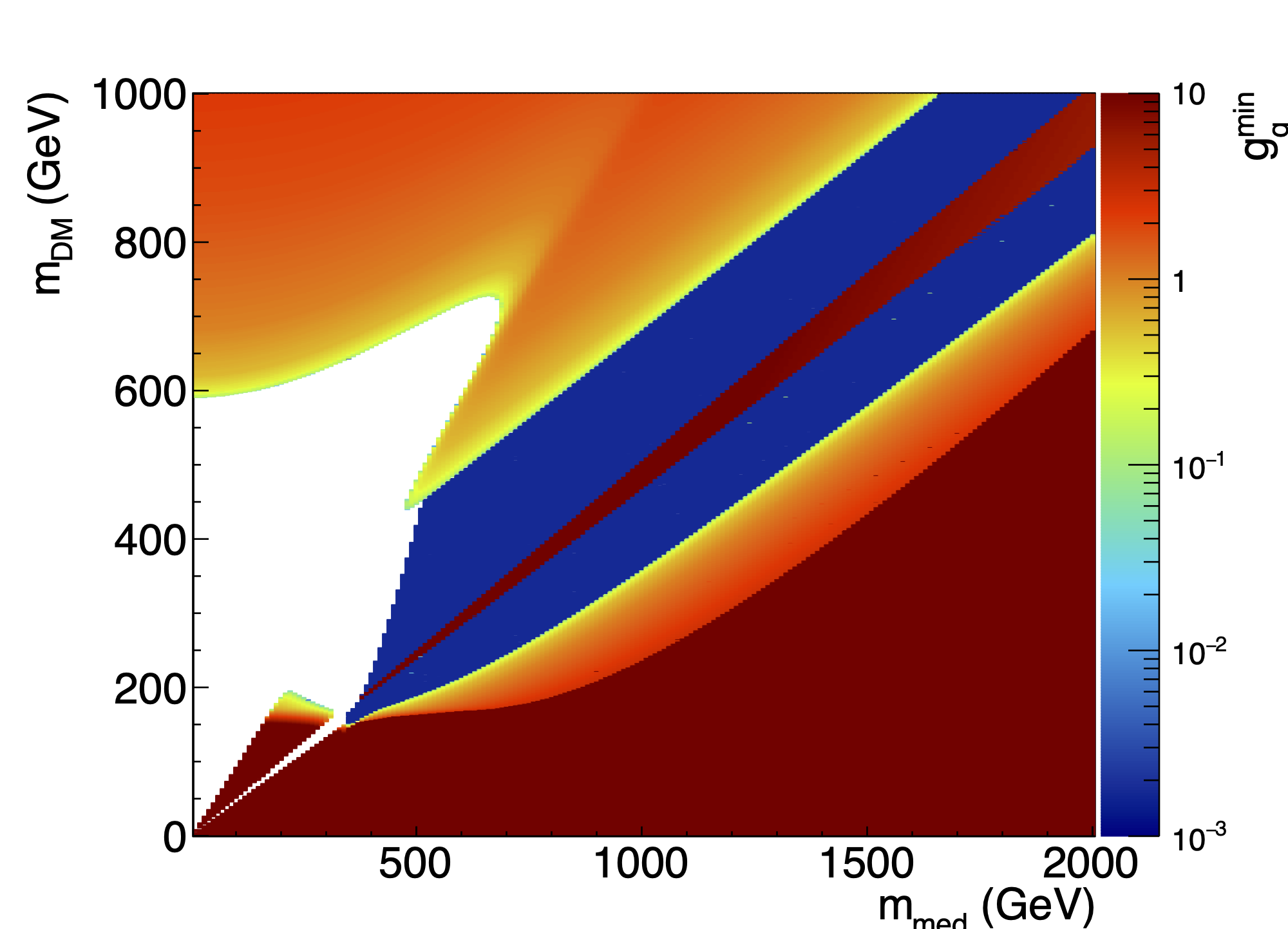}
    \caption{}
    \label{subfig:pseudoscalar}  
  \end{subfigure}
   \begin{subfigure}[b]{0.49\textwidth}  
    \includegraphics[width=\textwidth]{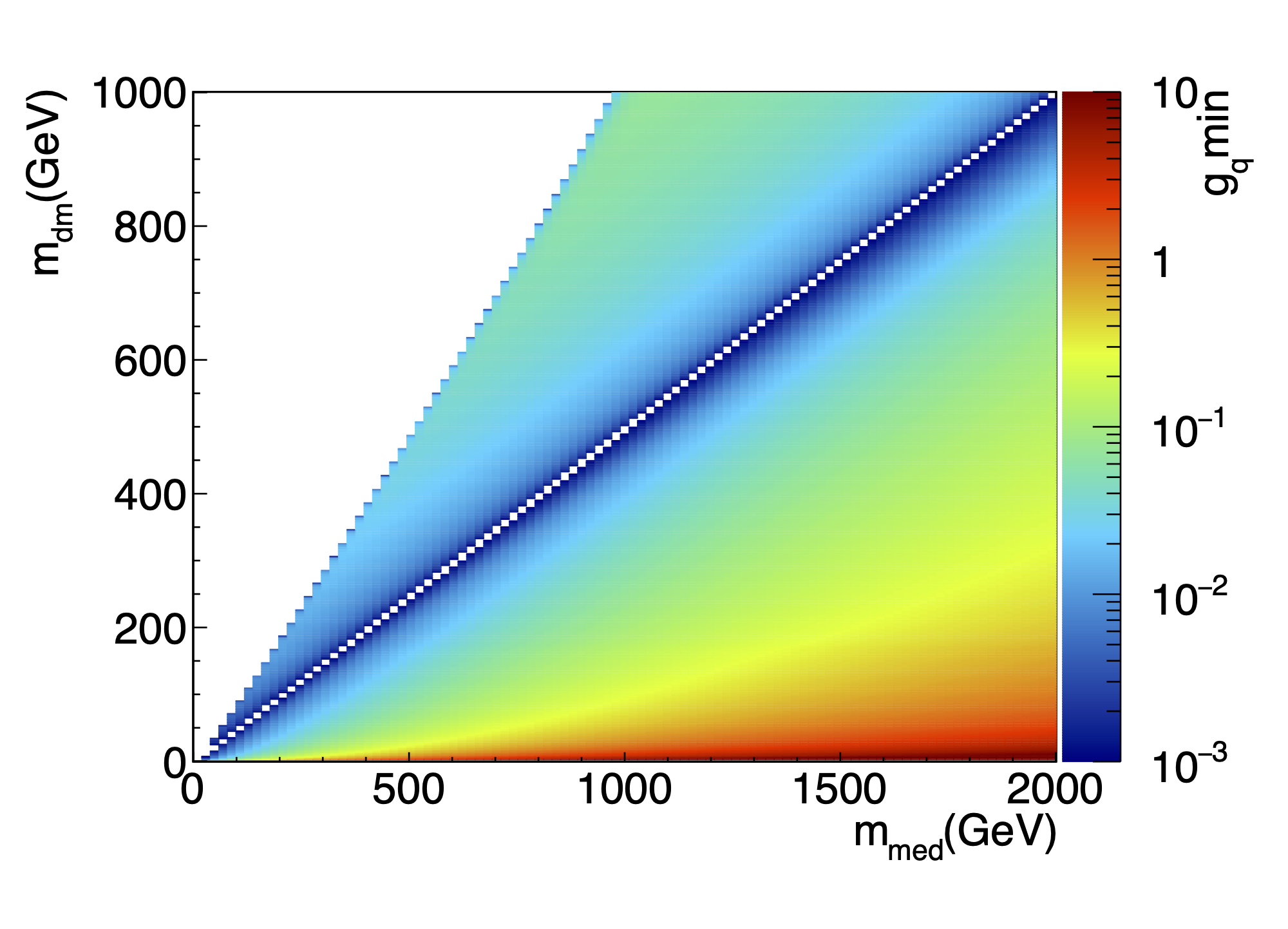}
    \caption{}
    \label{subfig:vector}
  \end{subfigure}
  \begin{subfigure}[b]{0.49\textwidth}  
    \includegraphics[width=\textwidth]{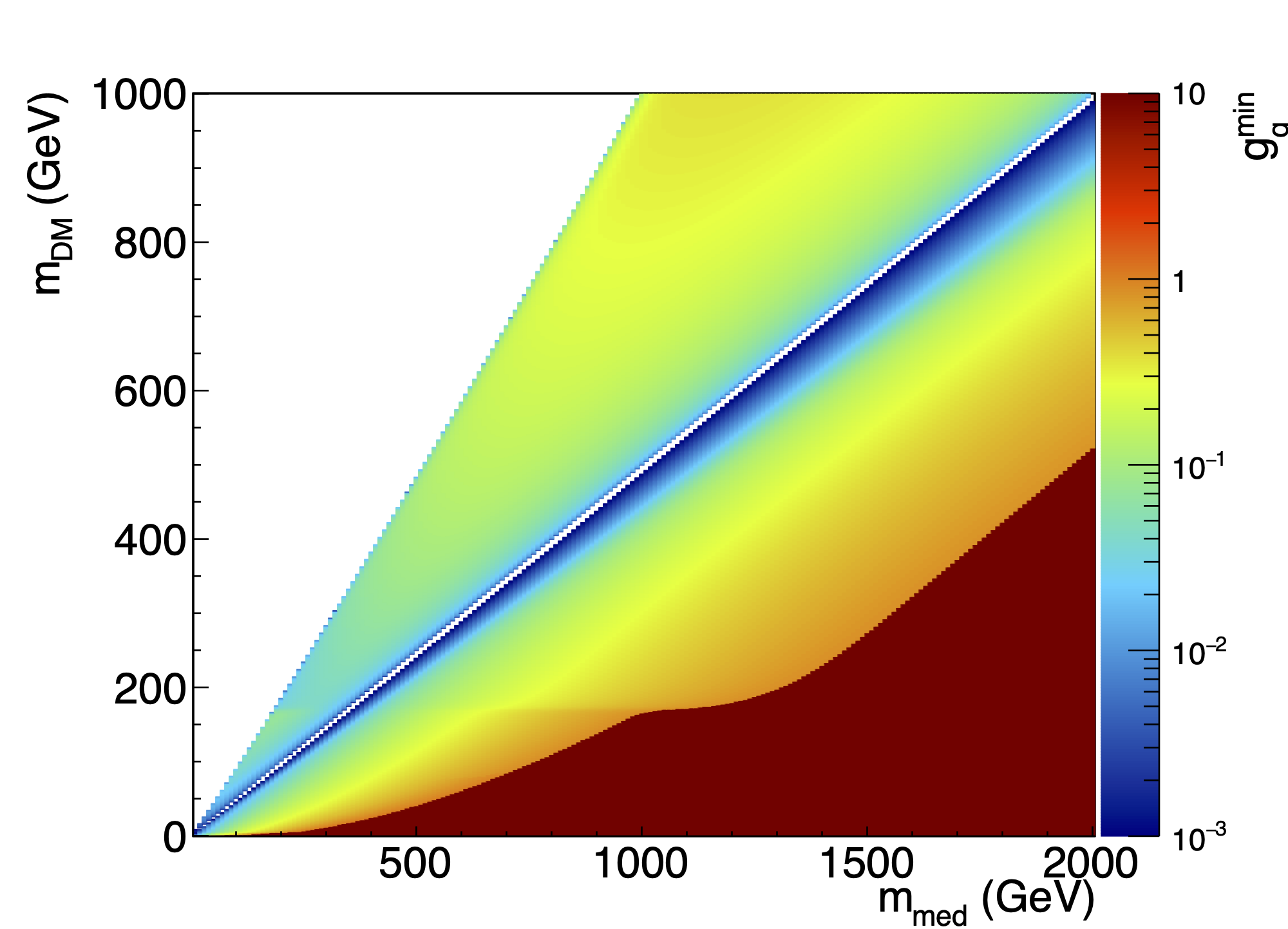}
    \caption{}
    \label{subfig:axialvector}  
  \end{subfigure}
  \caption{
    Minimum allowed coupling for the scalar (\subref{subfig:scalar}), pseudo-scalar(\subref{subfig:pseudoscalar}), vector(\subref{subfig:vector}), and axial-vector(\subref{subfig:axialvector}) mediators. The minimum coupling is computed on the z-axis. For the scalar and pseudo-scalar mediators, this coupling is treated as a correction to the Yukawa coupling.  
  }
  \label{fig:mincoupling}
  \end{center}
\end{figure}

\section{Conclusion}

The simplified dark matter models defined by the LHC Dark Matter Working Group provide interpretations covering a wide range of signatures and allowing multiple experiments to frame results in a single set of benchmarks.
However, the LHC experiments have historically been restricted in the range of these dark matter scenarios they can probe with resonant and \metplusx analyses due to the processing time required to generate Monte Carlo signal points in each scenario of interest. The methods described here for analytically rescaling between vector and axial-vector mediators as well as between different sets of couplings in the framework of the simplified DM models make this process significantly quicker by avoiding the need for computationally-expensive simulation.  The resonance analysis rescaling technique has been used in published CMS results for some time, but is publicly documented here for the first time by its developers. The \metplusx rescaling technique is new and, when combined with the resonance rescaling, makes it practical and time-efficient to reinterpret all major limits on the vector and axial-vector mediators, allowing experiments to create full limit summary plots in any scenario desired.

This paper acts as a user's guide to understanding and applying analytical rescaling methods to their own analysis limits. Certain caveats relating to the consistency of acceptances and $k$-factors across the rescaled scenarios have been described and recommendations outlined. The present version of this paper is intended to support the Snowmass~2021 process~\cite{snowmass21}.In a planned second version of this paper, a Python code package to handle the rescaling will be included. A preliminary version of this code is already available on Git, or through request to the authors, and has been passed to members of the ATLAS and CMS collaborations.
Although no rescaling formulae have so far been developed for lepton colliders or for scalar and pseudoscalar mediators, these are open points of investigation and may be added to the software in future if there is evident user interest.

This paper also provides a guide to the relic density implications of the LHC simplified models. The smallest possible couplings for which dark matter would not be overproduced are presented for all four simplified models. While these models are generally used more for their simplicity and ability to be directly compared across analyses and experiments more than for any probability as realistic dark matter models, their consequences for dark matter abundance are valuable to understand and provide an important additional axis of information for all users of the models. 


\acknowledgments 

Research by C. Doglioni is part of projects that have received funding from the European Research Council under the European Union’s Horizon 2020 research and innovation program (grant agreement 679305 and 101002463) and from the Swedish Research Council.
K.~Pachal's research is supported by TRIUMF, which receives federal funding via a contribution agreement with the National Research Council (NRC) of Canada.
Research for P.~Harris is supported by a Department of Energy Early Career Award.
Research by A.~Boveia is supported by the US Department of Energy (Grant DE-SC0011726).






\bibliography{CouplingScan-whitepaper}
\bibliographystyle{JHEP}

\end{document}